\documentclass[preprint, 3p, twocolumn]{elsarticle}

\usepackage{lineno}
\usepackage{ tabularx }
\usepackage{ graphicx }
\usepackage{ color }
\usepackage{ rotating }
\usepackage{ amssymb }
\usepackage{ siunitx }
\usepackage[markup=default]{ changes }
\usepackage{ xcolor,colortbl }
\usepackage{ todonotes }
\usepackage{url}
\usepackage{breakurl}
\usepackage[breaklinks]{hyperref}

\definecolor{Gray}{gray}{0.85}

\newcolumntype{a}{>{\columncolor{Gray}}c}
\newcolumntype{b}{>{\columncolor{white}}c}

\begin{document}
%\linenumbers
\begin{abstract}
The instrumental layout and technical realisation of the neutron resonant spin echo (NRSE) spectrometer RESEDA at the Heinz Maier-Leibnitz Zentrum (MLZ) in Garching, Germany, is presented. RESEDA is based on a longitudinal field configuration, boosting both dynamic range and maximum resolution of the spectrometer compared to the conventional transverse layout. The resonant spin echo technique enables the realisation of two complementary techniques: A longitudinal NRSE (LNRSE) option comparable to the classical neutron spin echo (NSE) method for highest energy resolution and large momentum transfers as well as a Modulation of Intensity with Zero Effort (MIEZE) option for depolarising samples or sample environments such as high magnetic fields, and strong incoherent scattering samples. With their outstanding dynamic range, exceeding nominally seven orders of magnitude, both options cover new fields for ultra-high resolution neutron spectroscopy in hard and soft condensed matter systems. In this paper the concept of RESEDA as well as the technical realisation along with reference measurements are reported.
 \end{abstract}


\begin{frontmatter}

\title{The longitudinal neutron resonant spin echo spectrometer RESEDA}

\author[TUM,MLZ]{C. Franz\corref{Cor}}
\ead{christian.franz@frm2.tum.de}

\cortext[Cor]{Corresponding author. +49 89 289 14760  }

\author[TUM,MLZ,TUD]{O. Soltwedel}

\author[MLZ]{C. Fuchs}

\author[MLZ,TUM]{S. S\"aubert}

\author[TUM,IAS]{F. Haslbeck}

\author[TUM]{A. Wendl}

\author[MLZ,BAY]{J. K. Jochum}

\author[TUM]{P. B\"oni}

\author[TUM]{C. Pfleiderer}

\address[TUM]{Physik Department, Technische Universit\"at M\"unchen, D-85748 Garching, Germany}
\address[MLZ]{Heinz Maier-Leibnitz Zentrum (MLZ), Technische Universit\"at M\"unchen, D-85748 Garching, Germany}
\address[BAY]{Bayerisches Geoinstitut, Universit\"at Bayreuth, D-95440 Bayreuth, Germany}
\address[IAS]{Institute for Advanced Study, Technische Universit\"at M\"unchen, D-85748 Garching, Germany}
\address[TUD]{Physik Department, Technische Universit\"at Darmstadt, D-64287 Darmstadt, Germany}

\date{\today}

\end{frontmatter}

%\maketitle

\section{Introduction}

Neutron Resonance Spin Echo (NRSE) \cite{1987Golub} is a Larmor labelling technique similar to Neutron Spin Echo (NSE) \cite{1972Mezei} that was developed to study amongst other things slow dynamics in liquid solutions, polymers, and spin glasses. In NRSE, traditionally employed in a transverse geometry (TNRSE, magnetic field perpendicular to the neutron flight path), the large solenoids generating the precession fields in NSE are replaced by pairs of much shorter resonant spin flippers (RSF). However due to the lack of sufficient manufacturing accuracy, TNRSE always stayed behind NSE by two orders of magnitude in resolution. But the technique proved to be very successful in combination with triple-axes spectrometry to resolve narrow phonon linewidths \cite{TRISP, TRISP1, 2015Groitl} and for Larmor diffraction \cite{1999Rekveldt}. One existing NRSE spectrometer for quasi-elastic scattering is located at LLB, France \cite{MUSES}.

The longitudinal NRSE (LNRSE) technique \cite{2005Haussler} overcomes the limitations of conventional transverse NRSE. First, the longitudinal rf flippers are by design self-correcting, which increases the highest resolution significantly. Second, the field subtraction method \cite{2016Krautloher} allows to reach infinitesimally small Fourier times, leading to a unique dynamic range, nominally exceeding seven orders of magnitude.
Furthermore the LNRSE method allows the use of Modulation of Intensity with Zero Effort (MIEZE) \cite{1997Hank, 1998Besenbock}, an NRSE variant that is insensitive to depolarising conditions such as large magnetic fields at the sample position and depolarising (e.g. ferromagnetic) samples. For strongly incoherent scattering samples MIEZE does not show the 2/3 background arising from spin flip scattering in contrast to conventional NSE (NRSE). Essentially, MIEZE is a high-resolution spin-echo TOF method using only the primary spectrometer arm for beam preparation \cite{1992Gaehler}. Typical applications comprise so far the investigation of quantum phase transitions, ferromagnetic materials, superconductors, skyrmions and hydrogenous samples. The MIEZE method is being actively developed at various places around the world, including the Reactor Institute Delft, the ISIS neutron source \cite{larmormieze} and JSNS at J-PARC \cite{2006Kawabata, 2013Hino}. Currently only BL06 at the J-PARC Materials and Life Science Experimental Facility (MLF) and RESEDA are in user operation with RESEDA being the only beamline worldwide based on longitudinal resonant spin flips.
In this paper we present the instrumental concept as well as the technical implementation of the LNRSE and MIEZE technique at the instrument RESEDA at the Heinz Maier-Leibnitz Zentrum (MLZ). Further we show the thorough characterisation of both measurement options, including resolution measurements and compare them to analytical calculations as well as neutron flux calculations using Monte-Carlo simulations. 

%------------------------------------------- Concept ---------------------------------------------

\section{Instrument concept}

The aim of the REsonance Spin Echo for Diverse Applications (RESEDA) spectrometer is to provide very high resolution neutron spectroscopy for various applications in hard and soft condensed matter.  Therefore RESEDA has been built to offer an extremely high dynamic range in time resolution, a broad band from small (SANS) to medium scattering angles as well as the ability to use a large variety of sample environments. To fulfill these requirements, an NRSE and a MIEZE option are implemented at RESEDA. Both share the primary spectrometer arm but make use of individual secondary spectrometer arms. A simultaneous operation is not possible, however switching from one mode to the other only requires a few minutes.  

\subsection{LNRSE}

The LNRSE option of RESEDA offers a resolution comparable to standard NSE spectrometers, outperforming transverse NRSE by two orders of magnitude in maximum Fourier time. The scattered beam size is restricted compared to NSE due to the use of the resonant flipper coils, limiting the detector size and hence solid angle coverage to 5.08\,cm (2\,inch) in diameter. The advantage over NSE is the increased dynamic range towards small Fourier times.
At RESEDA the LNRSE option is typically used for non-magnetic samples and without depolarising sample environment when highest resolution and/or large scattering angles are needed. Typical experiments include dynamics in bulk or confinement of macro molecules \cite{2011Sanz} or liquids \cite{2013Marry} and glass transitions in polymers \cite{2002Faivre}. For magnetic samples it is possible to separate nuclear and magnetic scattering by switching the $\pi$-flipper off. 

\begin{figure*}[htbp]
\includegraphics[width=0.99\linewidth]{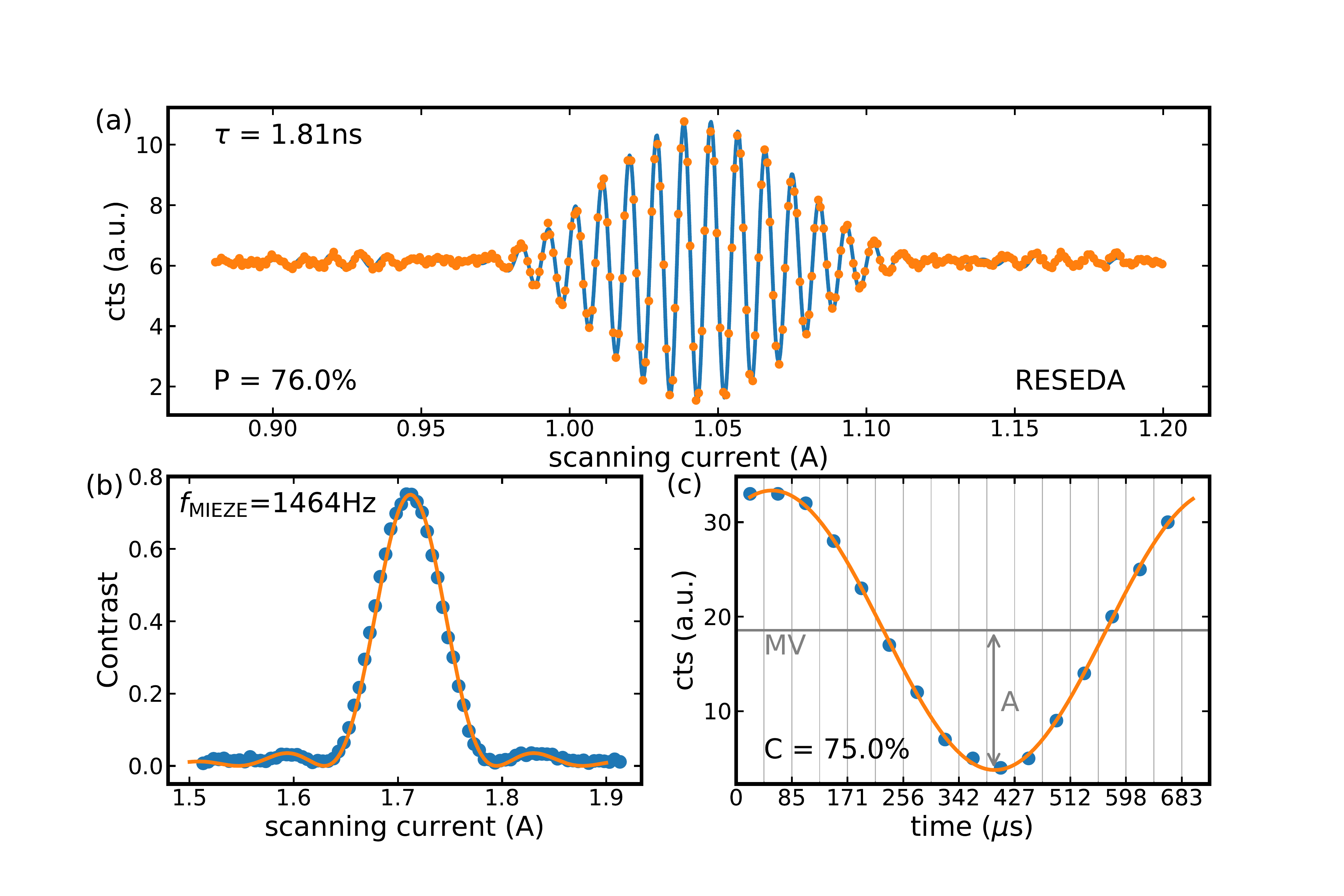}
\caption{Spin echo group in NRSE and MIEZE, respectively. (a) Spin echo group at a moderate Fourier time of 1.81\,ns in classical LNRSE mode. The intensity exhibits oscillations as a function of the current in the subtraction coil and the envelope is defined through the selector wavelength band. (b) The contrast of the MIEZE signal as a function of current in the subtraction coil. Here too, the shape of the MIEZE group is determined by the wavelength band. (c) Sinoidal intensity oscillations in MIEZE are in real time. The contrast shown in (b) is the ratio between the amplitude A and the mean value MV.}
\label{pic:segroups}
\end{figure*}

\begin{figure}[htbp]
\includegraphics[width=\linewidth]{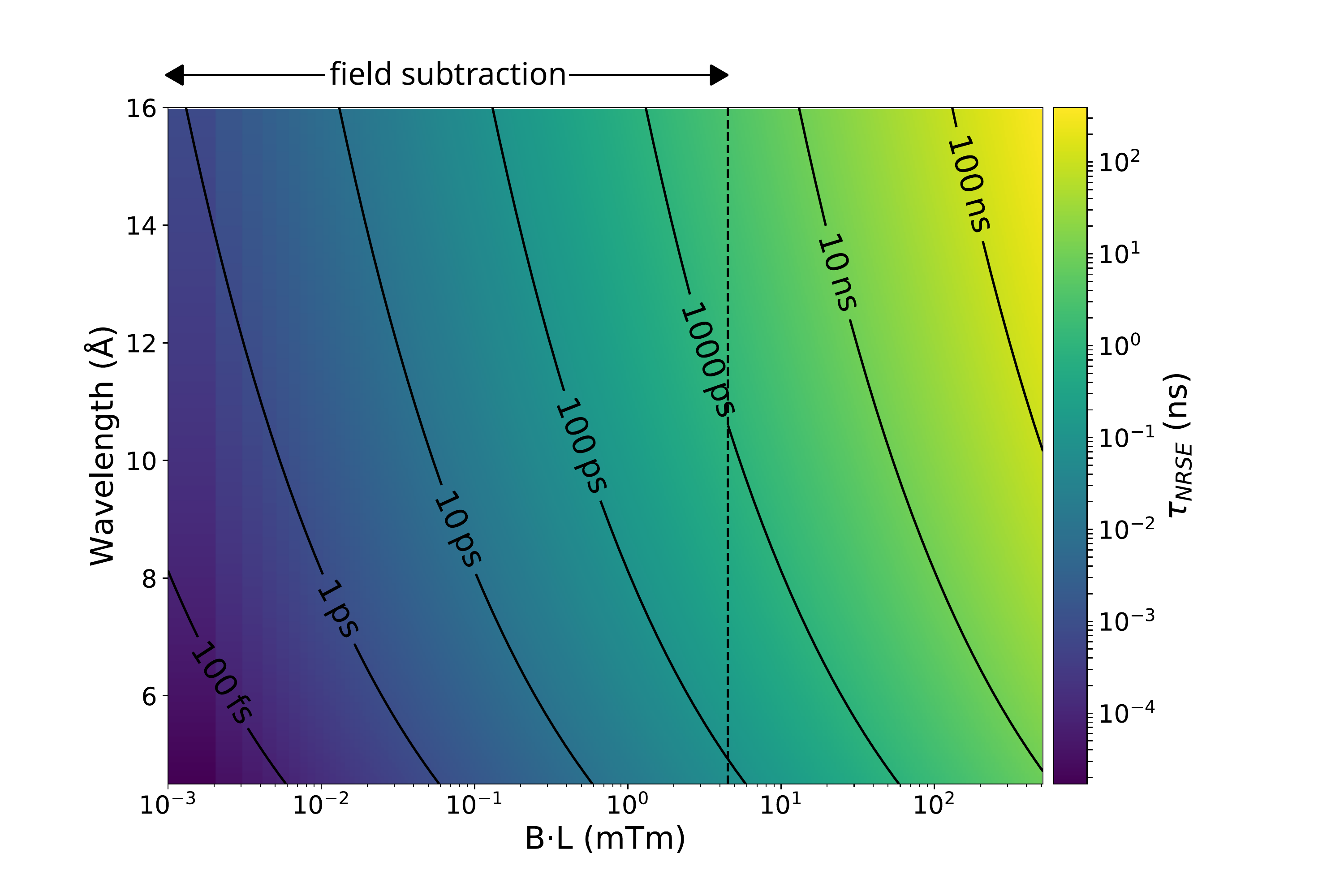}
\caption{Color coded RESEDA NRSE resolution depending on neutron wavelength and the field integral defined by the resonant flipper frequency and subtraction field. Contour lines denote the  Fourier time $\tau_{NRSE}$ in ns. The Fourier time depends linearly on the frequency, but has a cubic dependence on the neutron wavelength.}
\label{pic:res_n}
\end{figure}

In this mode two symmetric spectrometer arms are used, with a $\pi$-flipper before the sample position to reverse the precession direction. All four resonant coils are operated at the same frequency, with two coils per resonating circuit. To avoid frequency drifts between the spectrometer arms, frequency generators are coupled on a 10\,MHz clock signal. Small amplitude mismatches in the rf-coil-current, caused by manufacturing limitations, are compensated by built-in attenuators in the capacitance boxes. The NRSE signal is identical to an NSE echo with the envelope being the Fourier transform of the incident wavelength band and the oscillation frequency proportional to the mean wavelength (cf. Fig \ref{pic:segroups} (a)). During regular experimental operation only n points of the spin echo signal are recorded, with n typically ranging from 4 to 19. These are then fitted to yield the polarisation. The Fourier time can be calculated as 

\begin{equation}
    \tau_{\mathrm{NRSE}} = \frac{2 \gamma m_{\mathrm{N}}^2}{\hbar^2} \lambda^3 B \cdot L
\end{equation}

with $m_\mathrm{N}$ the neutron mass, $\lambda$ the neutron wavelength and the product of the static magnetic field $B$ and rf-flipper distance $L$. Note the factor two compared to NSE due to the resonant spin flip \cite{1987Golub}.
Fig. \ref{pic:res_n} shows the calculated resolution $\tau_{\mathrm{NRSE}}$ of the spectrometer color coded in ns depending on the neutron wavelength $\lambda$ in \AA\ and the field integral $B\cdot L$ in 10$^{-3}$\,Tm. For short Fourier times the frequency is kept at 35\,kHz due to the Bloch-Siegert shift \cite{1940Bloch} which determines the probability of a successful $\pi$-flip as the ratio of the static and rf-field strengths. In this regime the field subtraction method is used. Ramping up the magnetic field in the field subtraction coils (NSE coils) the field integral experienced by the neutrons may be reduced and short Fourier times are reached.

\subsection{LMIEZE}

In contrast to LNRSE, MIEZE is essentially a high resolution TOF method, which can be used for the investigation of magnetic materials or with depolarising sample environments such as large magnetic fields. It also proves to be very successful for the investigation of strong incoherent scattering samples, e.g. hydrogen containing materials. Using MIEZE the large background of 2/3 in conventional NSE originating from incoherent scattering can be avoided, as the information from the scattering event is encoded in the intensity modulation instead of the neutron Larmor phase. However the maximum resolution is reduced as compared to LNRSE. As for every TOF method, the sample size and shape are important to retain resolution \cite{2011Brandl, 2017Martin}.

\begin{figure}[htbp]
\includegraphics[width=\linewidth]{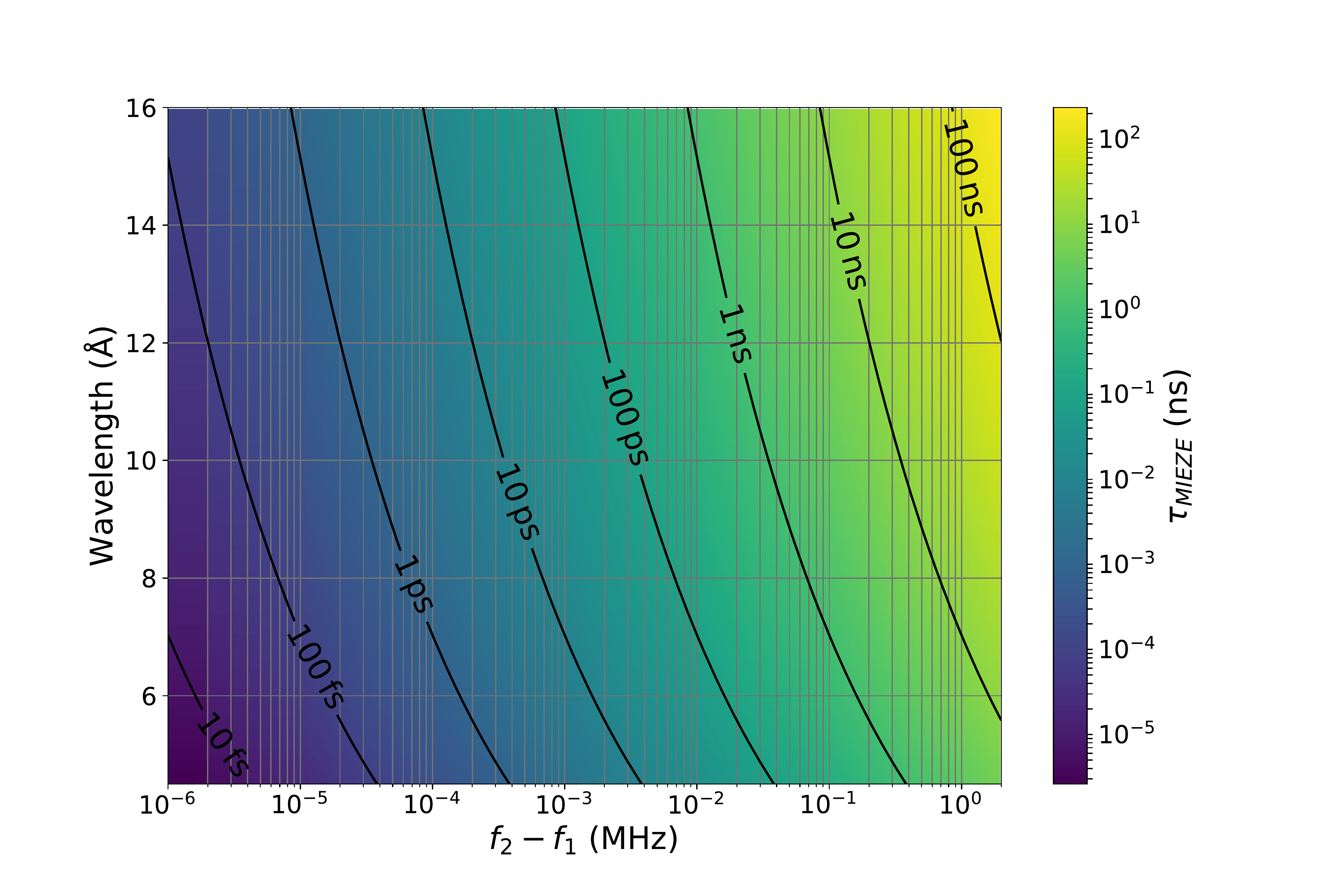}
\caption{Color coded MIEZE resolution available at RESEDA as a function of neutron wavelength and difference flipper frequency $f_2 - f_1$. Contour lines denotes Fourier time $\tau_{\mathrm{MIEZE}}$ in ns. The Fourier time depends linearly on the frequency, but to the power of 3 on the neutron wavelength.}
\label{pic:res_m}
\end{figure}

\subsubsection{Working principle}

In MIEZE mode only the primary spectrometer (in front of the sample) is used for beam preparation. Now both resonant flippers of the primary spectrometer arm are operated at different frequencies and the static magnetic fields at correspondingly different field values. The second flipper is operated at a higher frequency than the first flipper. This leads to a rotating polarisation for each wavelength after the second resonant flipper, which is subsequently transformed by a spin analyser into an intensity modulated signal \cite{1997Hank, 1998Besenbock}. The key advantage is that the analyser can be placed before the sample, making the method insensitive to magnetic fields at the sample position. Due to the use of a broad wavelength band the intensity modulation is only visible at the so called spin echo point as determined by the MIEZE condition

\begin{equation}
    \frac{f_b - f_a}{f_a} = \frac{L_1}{L_2}
\end{equation}

where f$_b$ and f$_a$ are the frequencies of the second and first flipper respectively, L$_1$ is the distance between the two resonant flippers and L$_2$ is the distance from the second flipper to the spin echo point, where the detector is placed.  The secondary spectrometer arm features an evacuated flight tube to minimize air scattering and a flat detector with fast readout. In this configuration the spin echo point can be moved in space (distance L$_2$) by varying the frequency ratio. At RESEDA a MIEZE ratio f$_b$/f$_a$ of 0.5 - 0.25 is used.

The Fourier time is then given by the difference in frequency $\Delta$f, the sample-detector distance L$_{sd}$ and the neutron wavelength:

\begin{equation}
    \tau_{\mathrm{MIEZE}} = \frac{2m_{\mathrm{N}}^2}{h^2} \cdot \Delta f \cdot L_{sd} \cdot \lambda^3
\end{equation}

Fig. \ref{pic:segroups} (b) shows the envelope of a MIEZE spin echo group in real space for a moderate modulation frequency of $\Delta$f\,= \,1464\,Hz. Resembling the envelope of the NRSE echo in Fig. \ref{pic:segroups} (a) even the side groups of the sinc$^2$ function are well reproduced. In contrast to NRSE the oscillations now appear in the time domain as shown in Fig. \ref{pic:segroups} (c). For every data point a fit over one period of the sinusoidally modulated signal in real time sampled with 16 data points is carried out. To gain sufficient statistics many periods of the sine are summed up. The polarisation in conventional NSE/NRSE is then replaced by the ratio of the amplitude A divided by the mean value of neutron counts MV:

\begin{equation}
    C = A/MV
\end{equation}

The so called contrast C denotes the ratio of high and low intensity, i.e. bright and dark detector pictures. A fine tuning of the echo point can be realised using the field subtraction coil, as the perfect ratio of the rf flipper frequencies is sometimes impossible due to the lack of resonances at the desired frequency. As in NRSE and NSE, the width of the envelope depends on the wavelength spread of the incoming neutron beam.

For the measurement of inelastic signals, e.g. the spin wave dispersion in ferromagnetic iron \cite{2019Saeubert} the precise control of the phase shift of the intensity modulated beam between sample and reference measurement is of crucial importance. To achieve this, all frequency generators of both resonance circuits as well as the detector trigger and time chopper are running on a single quartz (common 10\,MHz clock signal).

\begin{figure}[htbp]
\centering
\includegraphics[width=0.8\linewidth]{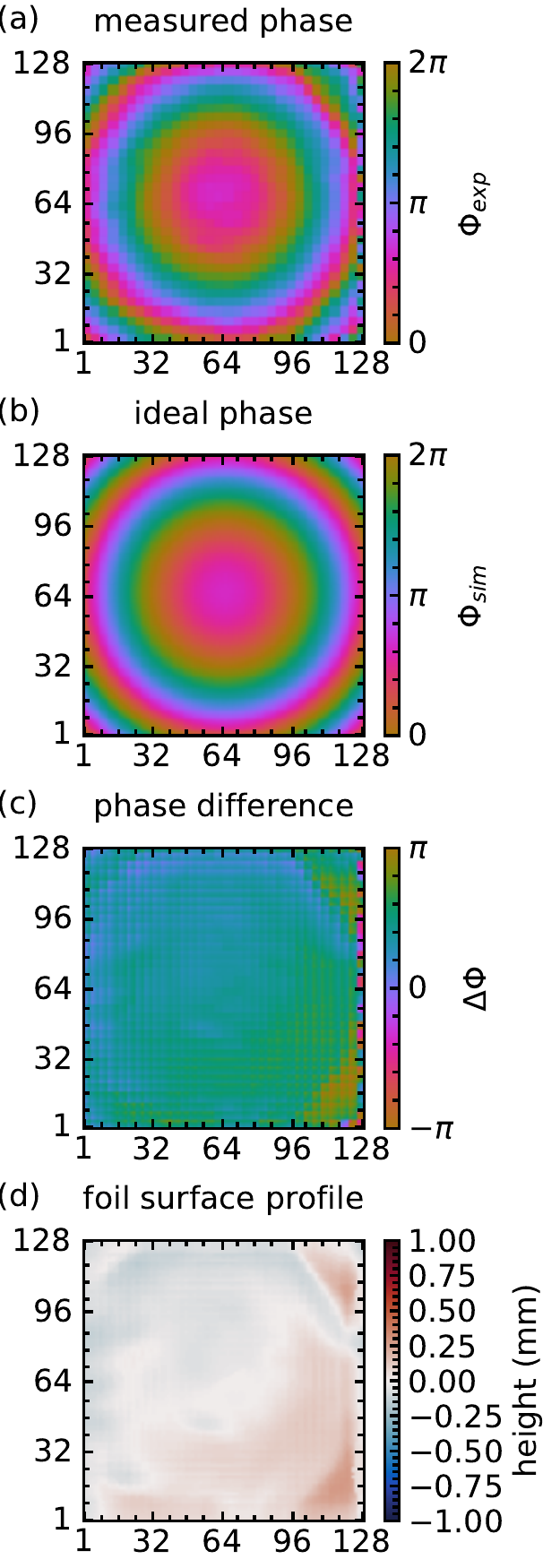}
\caption{Phase rings on the second foil of the 2D CASCADE psd detector derived from an elastic scatterer. The second foil of the detector was used with instrument settings $\lambda$\,=\,6\,\AA, $L_{\mathrm{SD}}$\,=\,3.43\,m and $f_{\text{MIEZE}}$\,=\,342\,kHz. A phase shift of more than 2$\pi$ is observed over the detector area of 20\,x\,20\,cm$^2$. (a) ideal phase derived from analytical calculations (b) Measured phase (c) Phase difference between (a) and (b). (d) calulated foil topography derived from (c)}
\label{pic:phase}
\end{figure}

%---------------------------------------------------------------------------------------

\subsubsection{Detector requirements}

As the spatial width of the MIEZE group is shrinking with increasing Fourier time and the frequency of the intensity oscillations increases, a fast and flat detector is required. At RESEDA a bespoke 2D position sensitive CASCADE detector from CDT Heidelberg, Germany with an active area of 20\,x\,20\,cm and a spacial resolution of 128\,x\,128\,px and 6 foils is used. The single conversion layers are $\sim$\,1$\mu$m thick and the maximum readout per timechannel is 100\,ns (10\,MHz). Whenever the width of the MIEZE group allows, all 6 foils can be phase corrected and combined to reach an efficiency of $\sim$\,60\,\% at a neutron wavelength of 6\,\AA. 

For long Fourier times ($\tau_{\mathrm{MIEZE}}$\,$>$\,100\,ps) phase rings around the direct beam occur on every foil due to individual flight path lengths for neutrons from the sample to the detector. These phase rings (cf. Fig \ref{pic:phase} (a)) must be corrected when the region of interest on the detector contains pixels with different phases, otherwise the contrast is smeared out in this region. This phase correction is typically done using a resolution measurement with an elastic scatterer, where the whole detector is illuminated. From this measurement a phase map for every foil is created (Fig. \ref{pic:phase} (b)), which is then used to shift the sample data accordingly. A software for data reduction including all corrections is currently under development and will be published elsewhere together with a more detailed description of the data reduction process \cite{2018data}. 

%-----------------------------------------------------------------------

\subsubsection{Sample geometry limitations}

As for every time-of-flight method the maximum achievable contrast in MIEZE depends on the scattering angle and sample geometry. Scattered neutrons from different parts of the sample can reach the same detector pixel with different flight path lengths. If these flight paths and therefore flight time differences reach the order of one period of the sinusoidal intensity modulation, the signal is damped even for elastic scattering, making a measurement impossible. 

\begin{figure}[htbp]
\includegraphics[width=\linewidth]{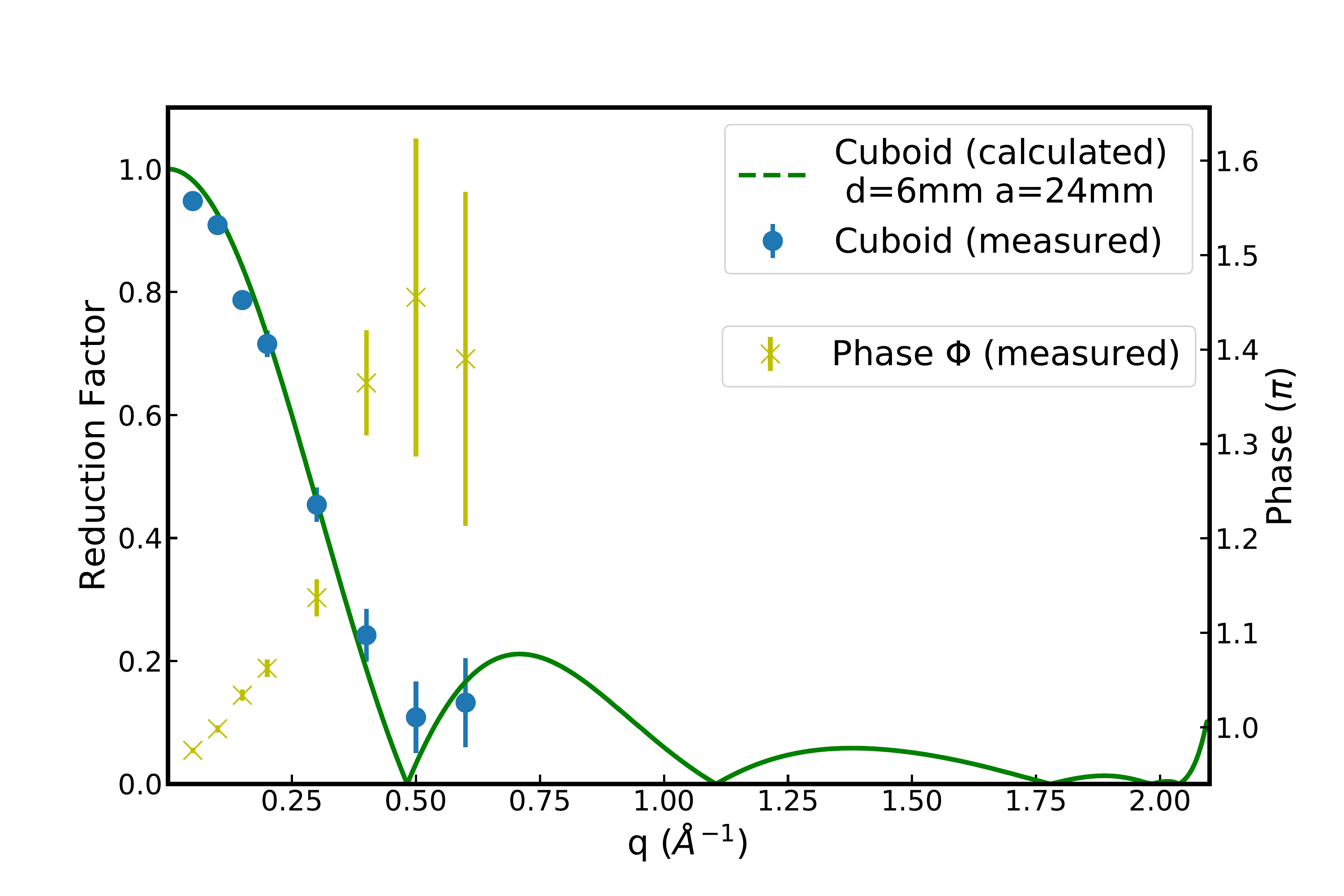}
\caption{MIEZE reduction factor (or maximum achievable contrast) as a function of the scattering vector for a elastic graphite sample in a defined geometry and a fixed Fourier time. The solid line is the calculated reduction factor; the data points are the reduction factor measured on RESEDA; the crosses correspond to the measured phase in radiant.}
\label{pic:reduction}
\end{figure}

This reduction factor can be calculated for a given wavelength and Fourier time, taking the exact sample geometries into account  \cite{2011Brandl, 2017Martin}. Fig. \ref{pic:reduction} shows the calculated reduction factor for a cuboid sample of 6\,mm thickness and 24\,mm width as a function of the scattering vector along with data measured on RESEDA for such a sample. A neutron wavelength of 6\,\AA\ and a fixed Fourier time of 500\,ps is assumed. A negative reduction factor can be understood as a phase shift in the signal, as most scattered neutrons experience a flight path difference accounting for a $\pi$-shift. Using the phase lock and comparing the phase to a direct beam measurement, this phase shift can be observed on RESEDA. A tool to calculate the resolution and scattering angle dependent reduction of the signal strength for various sample geometries is available for users on the RESEDA website \cite{redfac}.

%--------------------------------------------------Technical realisation -------------------------------------------------------

\section{Technical realisation}

RESEDA is located at the end station of the curved (radius 1640\,m) neutron super mirror guide (m\,=\,2) NL5-S (length 70\,m, initial cross section 170 x 29\,mm$^2$), and uses the upper third of this guide, a section measuring 36\,x\,36\,mm$^2$. The guide is shared with the reflectometer TREFF. It is fed by cold moderated neutrons from the FRM\,II research reactor which possess a Maxwellian spectrum with a usable neutron flux between 3.5\,$\text{\AA}$ and 20\,$\text{\AA}$ \cite{2006Zeitelhack}. In order to get monochromatic neutrons and keep the neutron wavelength as well as its spread flexible, a velocity selector (3h, Astrium NVS-032, Airbus DS GmbH, Germany) on a circle segment (1-Circle Segment 5202.80, HUBER Diffraktionstechnik GmbH Co. KG, Germany) with a maximum angular speed of 28300\,rpm is implemented.

\begin{figure}[hbtp]
\includegraphics[width=\linewidth]{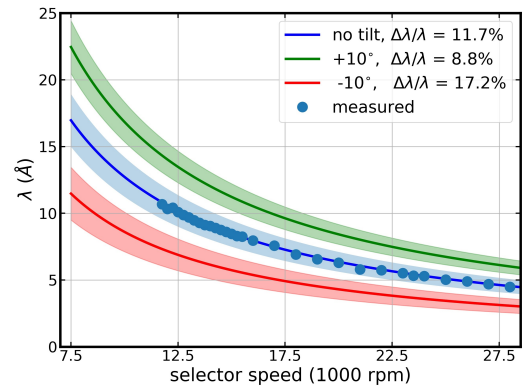}
\caption{Wavelength versus selector rotation speed at \ang{-10}, \ang{0} and \ang{+10} tilt angle of the neutron wavelength selector. The corresponding wavelength band is given in transparent color. Measurement of the wavelength is done by evaluation of the spin echo group.}
\label{pic:nvs32}
\end{figure}

In this combination wavelengths from 3.5 to 20\,$\text{\AA}$ with a spread of 8.8 to 17.2\,\% are available and it is thus possible to  offer a high flux / relaxed resolution as well as a reduced flux / high resolution option (c.f. Fig. \ref{pic:nvs32}). To attenuate the high flux of the direct beam on the detector three electro-pneumatic absorber plates can be placed in the beam. All of them are made of 5\,mm thick borated plastic and they completely cover the neutron beam path. To make them partially transparent for neutrons they have isotropically drilled holes of varying density. Thus neutron wavelength independent attenuation factors of 3, 15 and 30 can be arbitrarily combined to gain an optimal neutron flux. 

\begin{figure*}[htbp]
\includegraphics[width=\linewidth]{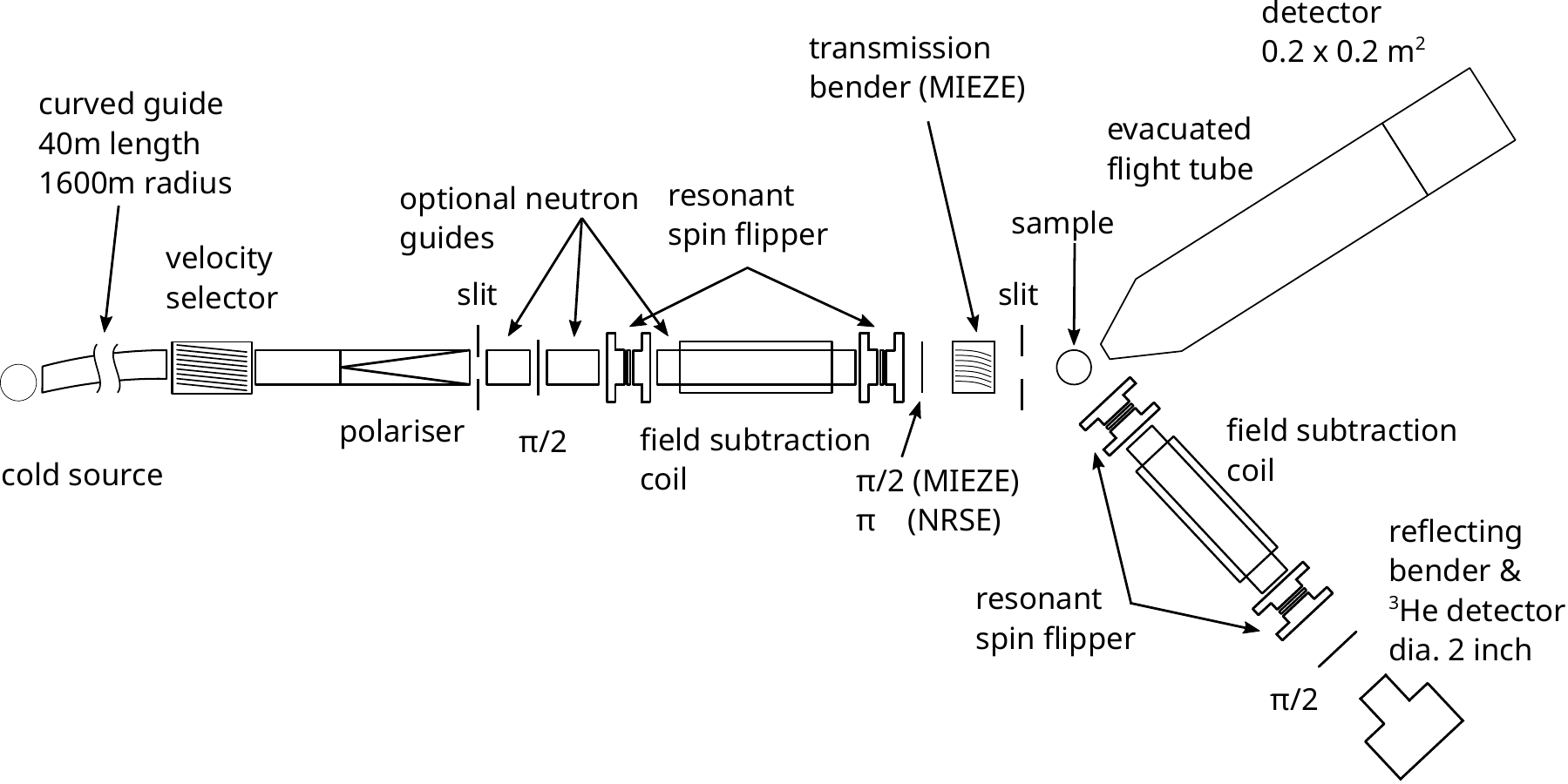}
\caption{Schematic depiction of the RESEDA instrument layout at the beam position NL5-S at FRM\,II (MLZ). Two secondary spectrometer arms for the NRSE and MIEZE options are available.}
\label{pic:schematic}
\end{figure*}

\subsection{Primary spectrometer}

A 2\,m long double V-cavity equipped with m\,=\,5 FeSi transmission supermirrors is used to polarise neutrons within a wavelength range from 3.5 to 15\,$\text{\AA}$ with a transmission and polarization \textgreater\,98\%. The V is oriented such that it points towards the neutron source with vertical supermirrors. Stacks of NdFeB magnets and steel plates enclose the cavity and act as a magnetic cage, providing a vertical field of 50\,mT to keep the multilayers of the cavity in magnetic saturation. For larger wavelengths and divergent beams a second, interchangeable V-cavity, covering the wavelength band from 8 to 25\,$\text{\AA}$, is currently under review.  Downstream of the polariser a non-magnetic motorized cross slit (JJ X-ray IB-C50-AIR) is installed. Precise mechanics, including encoders on all 4 axes and custom made edged B$_4$C blades, allow reproducible shaping of the primary beam while minimizing stray scattering. An equivalent slit system is installed approximately 4\,m further downstream in front of the sample position (the exact position may vary due to experimental conditions such as sample environment). Using both slits, it is possible to adjust the beam size as well as beam position to the shape and location of the sample. The beam divergence can also be tuned with this slit system. Right downstream of the first slit the beam monitor is positioned. This $^3$He based counter (MNH10/4.2S, former Dextray, later Canberra) has an efficiency of 1:10$^5$ at 1.5\,$\text{\AA}$ while operated at 1\,kV. A MRS2000 (Mesytec, Germany) pre-amplifies the signal before it is registered in a PC-counting card. \\
The primary spectrometer arm is equipped with m\,=\,1.2 straight neutron guides, interrupted only by the static spin flippers and resonant flipper coils. The neutron guides yield a factor of five in intensity at the sample position, if the experiment allows for a high divergence in the beam. For experiments at very low q, a set of four pinholes with openings from 1\,mm to 20\,mm can be installed manually and act as a collimation track. 
Polarization of the neutron spins is guaranteed by a small longitudinal guide field of several Gauss throughout the whole instrument, generated by a set of in total 20 coils with a diameter of 40\,cm and 60 windings of copper wire. These coils are additionally used to perform the adiabatic transitions from the transverse field of the polariser or analyser into the neutron flight path.

\subsection{Resonating circuits}

The very heart of a LNRSE spectrometer are the resonant spin flippers, a combination of a static magnetic field in neutron flight direction and a fast oscillating magnetic field, perpendicular to the neutron flight path, linked by the relation

\begin{equation}
    \omega_L = \gamma \cdot B
\end{equation}

with $\omega_L$ the Larmor precession frequency of the neutron, $\gamma$ the gyromagnetic ratio and $B$ the static magnetic field \cite{1987Gaehler,1987Golub}. For a maximum frequency of 4\, MHz a static magnetic field of B\,=\,0.137\,T is needed, yielding a field integral of 0.51\,Tm with a flipper distance of 1.87\,m. The ac-current of up to 4.4\,A (at $\lambda_{min}$\,=\,4.5\AA) in the resonant flipper coils is generated by a resonating circuit consisting of a frequency generator (Agilent 33250a), an amplifier (Rhode \& Schwarz BBA150), and bespoke in-house developed capacitance boxes for impedance matching, filtering of the signal and tuning to the desired frequency by adjusting the capacitance. 

The spin flip coils are standard state of the art rf coils \cite{2015Groitl} adjusted for operation at high frequencies. Read-back and monitoring of the oscillation through individual pick-up coils for every resonant coil is done with a Rhode \& Schwarz RTE1104 4-channel digital oscilloscope with 16 bit vertical resolution. The lower frequency limit is 35\,kHz due to the Bloch-Siegert shift \cite{1940Bloch}. 

\subsection{Field subtraction coils}

To extend the dynamic range of the spectrometer towards smaller Fourier times (higher energy transfers) than given by the minimum frequency of 35\,kHz at the lowest neutron wavelength, a field integral subtraction method is used \cite{2016Krautloher}. Therein a conventional NSE coil (l\,=\,1\,m), placed between the rf flippers, is used to reduce the field integral. Since elongated coils suffer from inhomgeneities, a cos$^2$ geometry was chosen to optimize the generated field shape \cite{1996Zeyen}. The field integral generated by the NSE coils is 5.20\,$\cdot$\,10$^3$\,Tm at 2\,A, larger than the field integral of 4.48\,$\cdot$\,10$^3$\,Tm generated at $f$\,=\,35\,kHz, allowing for maximum dynamic range. The coils are operated by a Caenels FAST-PS 0580-400 fast switching power supply with a 18\,bit current setting resolution.

\subsection{Sample environment}

The sample table from HUBER Diffraktionstechnik GmbH features a rotation stage, x- and y-translation as well as x- and y-cradle, usable with standard FRMII sample environment (cryostats, furnaces, small magnets \cite{FrmIISE}). 
For large magnets (e.g. horizontal 5\,T SANS, 12\,T vertical magnet) only the rotation stage is usable. A larger sample table with all necessary degrees of freedom  for sample environment up to 900\,kg is in the planning.
The available sample environment offers temperatures from 50\,mK (dilution inserts) to 2150\,K (induction furnaces). Additionally within the MIEZE technique magnets with fields up to 12\,T vertical or 17\,T horizontal can be used \cite{2015Kindervater}. Special requirements like moisture cells or the application of electric fields are possible.

\subsection{NRSE secondary spectrometer}

The begin and the end of the neutron spin precession zone in LNRSE mode is determined by Mezei-type $\pi$/2 flippers after the polariser and before the analyser. In front of the sample table a $\pi$ flipper reverses the precession. The secondary spectrometer arm is identical in construction to the primary arm with two resonant spin flippers and one subtraction coil in between. Small deviations in the RSF flipper distance to the primary arm can be balanced by the subtraction coil current. An additionally installed phase shifting coil around the subtraction coil allows precise tuning of the spin echo group. 
Neutron analysis and detection in LNRSE mode is done by a reflecting bender (P\,$>$\,0.9 for 3\,\AA\,$<$\,$\lambda$\,$<$\,15\,\AA) and a standard $^3$He end-window counter tube (Eurisys 15NH5/5X, 2\,inch diameter) with Mesytec MRS2000 pre-amplifier.

\subsection{MIEZE secondary spectrometer}

In MIEZE mode, the $\pi$ flipper in front of the sample environment described earlier is used as a $\pi/2$ flipper to stop the precession of the neutron spin. After the $\pi/2$ flipper, an adiabatic transition from the longitudinal guide field into the transverse field of the analyser is performed. For the MIEZE option a transmission bender from SwissNeutronics followed by \ang{;120;} vertical collimation to absorb reflected neutrons is placed between the second $\pi$/2 flipper and the sample position. The so-created fast oscillating neutron intensity with frequencies of up to several 100\,kHz is recorded by a flat and fast $^{10}$B based 2D-CASCADE detector from CDT Heidelberg GmbH \cite{2011Haeussler, 2016Koehli}. To increase efficiency 6 foils with an active area of 20\,x\,20\,cm and a 1\,$\mu$m conversion layer each are stacked in the detector, yielding an efficiency of 50\% at 5.4\,\AA. A 10 foil CASCADE detector with increased efficiency is currently under commissioning. From the sample position to the CASCADE detector an evacuated flight tube with bespoke nose for different sample environments prevents air scattering and absorption of neutrons. The last 0.5\,m of the flight tube are filled with He, as the detector front window is too thin to be directly pumped.

\begin{table}
\begin{tabular}{l|l}
\rowcolor{Gray}
\multicolumn{2}{c}{Neutron Guide system}\\
    Neutron guide   &  NLS5-S \\
    cross section   &  36\,x\,36\,mm$^2$ \\
    coating         &   m\,=\,2\\
\\
\rowcolor{Gray}
\multicolumn{2}{c}{Wavelength selector}\\
    wavelength      &   3.5..22$\text{\AA}$ \\
    tilt            &  -10..10$^{\circ}$ \\
    wavelength band &   8.8..17.2\% \\
\\
\rowcolor{Gray}
\multicolumn{2}{c}{Collimation system}\\
    double slit system  & 1\,x\,1\,..\,50\,x\,50\,mm$^2$ \\
    solid state collimators & \ang{;80;}, \ang{;40;}, \ang{;20;}\\
\\
\rowcolor{Gray}
\multicolumn{2}{c}{Polarization}\\
    Type  & Double V-cavity \\
    coating & FeSi, m=5 \\
    length & 2\,m \\
    taper angle & 1.04$^{\circ}$ \\
    wavelength & 3.5..15$\text{\AA}$ \\
    polarization & \textgreater\,98\%\\
\\ 
\rowcolor{Gray}
\multicolumn{2}{c}{Polarisation analysis}\\
    NRSE    & reflection bender \\
    MIEZE   & transmission bender \\
            & 5-fold V-cavity \\
\\
\rowcolor{Gray}
\multicolumn{2}{c}{Q-range}\\
    NRSE    & 0.03..2$\text{\AA}^{-1}$ \\
    MIEZE   & 0.01..1$\text{\AA}^{-1}$ \\
\\
\rowcolor{Gray}
\multicolumn{2}{c}{Detector system}\\
    NRSE            & $^3$He one inch diameter \\
    MIEZE           & 2D $^{10}$B CASCADE \\
    size            & 20\,x\,20\,cm\\
    pixels          & 128\,x\,128\\
    time resolution & 100\,ns (10\,MHz)\\
\\
\rowcolor{Gray}
\multicolumn{2}{c}{Resolution}\\
    NRSE    & 0.001..20\,ns \\
    MIEZE   & 0.0001..10\,ns \\   
\\
\rowcolor{Gray}
\multicolumn{2}{c}{Flux}\\
    with guides     & 2\,$\cdot$\,10$^7$\,n/cm$^2$s (at 6\,$\text{\AA}$)\\
    w/o guides      & 0.5\,$\cdot$\,10$^7$\,n/cm$^2$s (at 6\,$\text{\AA}$)
\end{tabular}
\caption{Summary of the technical data for the neutron resonance spectrometer RESEDA}
\label{tab:tech_data}
\end{table}

%-------------------------------------------Performance----------------------------------------------------------

\section{Instrument performance}

\subsection{Neutron flux}
Neutron flux at the sample position of RESEDA was calculated using McStas 2.5 and verified by measurements with a calibrated fission chamber. McStas simulations were carried out using the cold source from \cite{2006Zeitelhack}, the curved NL5-S guide system, velocity selector and the double V-cavity polariser. Neutron guides on the primary spectrometer arm were taken into account whereas the resonant and static spin flippers were accounted for by gaps in the neutron guides. Contrary to transverse NRSE the spin flippers in LNRSE expose only a few milimeters of aluminum to the beam and do not attenuate it. An inhomogeneity of the beam profile due to the wide curvature of the neutron guide was not observed in simulations. 

\begin{figure}[htbp]
\includegraphics[width=\linewidth]{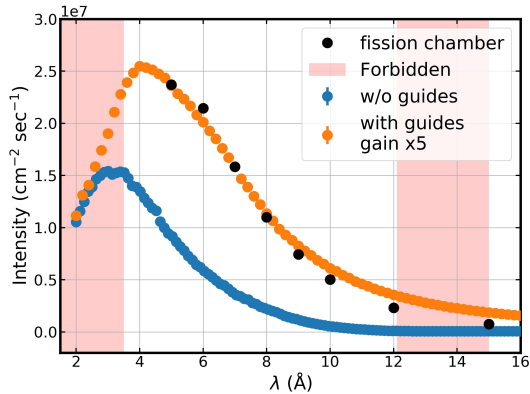}
\caption{Neutron flux in cm$^{-2}$sec$^{-1}$ versus neutron wavelength at the sample position from McStas simulations with m\,=\,1.2 neutron guide on the primary spectrometer arm (blue) and without (orange). The forbidden wavelength ranges by selector vibrations are shaded in red.}
\label{pic:flux}
\end{figure}

Results of the simulations are shown in Fig. \ref{pic:flux} with and without the use of neutron guides. The red shaded areas mark forbidden wavelength ranges due to vibrations of the neutron wavelength selector at zero tilt angle. Whenever the experiment allows, a factor of five can be gained in neutron flux by using neutron guides. The maximum intensity is then shifted from 3.5\,\AA\ to 4.5\,\AA. The flux at longer wavelengths also gains from the neutron guides. There is no influence of the neutron guides on the contrast for Fourier times below 1\,ns.
Simulations were checked by measurements with a calibrated $^{235}$U fission chamber with an efficiency of 1.85$\cdot$10$^{-4}$ at 4.751\,\AA. A 6\,mm pinhole was used as an entrance slit, defining the detector area. 

The overall wavelength dependence is reproduced very well, but with an overestimation of neutron flux at wavelengths larger than \,8\,\AA. This behaviour is well known for the simulated cold source of FRM\,II \cite{Ostermann}. Quantitatively the simulation overestimates the neutron flux by a factor of two. This discrepancy is currently of unknown origin.  Note that for MIEZE measurements the additional analyser is reducing the flux at the sample position.

\subsection{MIEZE}

\begin{figure}[htbp]
\includegraphics[width=\linewidth]{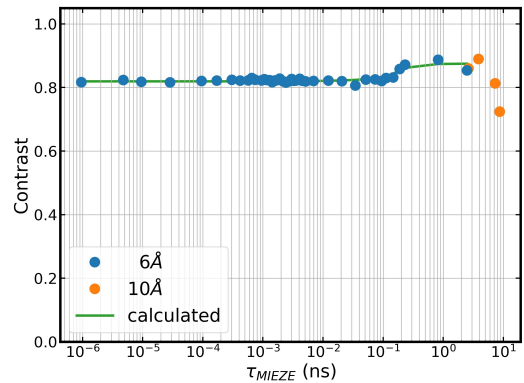}
\caption{Experimental MIEZE resolution for neutron wavelengths of 6\,\AA\ and 10\,\AA\ over 7 orders in Fourier time. The solid line is an analytical calculation considering the Bloch-Siegert shift.}
\label{pic:res_mieze}
\end{figure}

Fig \ref{pic:res_mieze} shows a typical MIEZE resolution function measured on an elastic scatterer (graphite, 20\,mm pinhole as sample slit) for wavelengths of 6\,\AA\ and 10\,\AA, respectively over 7 orders of magnitude in Fourier time. The data has been taken for small scattering angles, where the reduction factor is neglectable. Along with the experimental data a solid line representing an analytical calculation considering the Bloch-Siegert shift \cite{1940Bloch} is shown. The maximum polarisation after each resonant flipper is slightly damped at low frequencies, leading to a reduction of the overall contrast in the field integral subtraction region \cite{2019Franz} where both resonant flipper coils are operated at or close to 35\,kHz. The calculated curve describes the experimental data in detail, indicating the good control of the technique. The last two data points at 7.32\,ns and 8.77\,ns recorded at 10\,\AA\ show a drop in contrast, however the signal is stable and reproducible. The reason for this drop is unclear and under further investigation. 

%\subsection{NRSE}

%\begin{itemize}
%    \item Resolution curve NRSE for different wavelength (to be measured)
%\end{itemize}

%\begin{figure}[htbp]
%\includegraphics[width=\linewidth]{NRSE_reso.png}
%\caption{Experimental NRSE resolution. To be measured...}
%\label{pic:res_nrse}
%\end{figure}

%--------------------------------------------------------------------------------------------

\section{Conclusions}

In this paper we have given a summary of the current status of the technical setup of the resonant spin echo spectrometer with MIEZE option RESEDA at MLZ. We have shown the achievable resolution for both measurement options, NRSE and MIEZE. The NRSE option is dedicated for highest resolution and large scattering angles, while the MIEZE option allows for depolarising samples and sample environments as well as strong incoherent scatterers at small and intermediate scattering angles.

The combination of MIEZE and a 17\,T unshielded longitudinal magnet has already been successfully tested \cite{2015Kindervater}. As a time-of-flight method the maximum achievable resolution is coupled to the sample geometry and scattering angle, which can be calculated prior to the experiment. Both options strongly benefit from the field subtraction method being unique to longitudinal resonance spin echo, increasing the dynamic range drastically. Taken together both options make RESEDA a versatile instrument for the investigation of fast and slow dynamics at small or wide angles in soft and hard condensed matter. Showcase experiments including superconducting ferromagnets \cite{2019Haslbeck}, spin ice systems \cite{hto}, skyrmions \cite{JPSJ} and water \cite{water} have been completed and demonstrated the excellent performance of RESEDA. The NRSE option is currently under commissioning and first results will be published along with a detailed experimental resolution elsewhere.

%---------------------------------------------------------------------------------------------------------------

\section{Acknowledgements}

We wish to thank A. Mantwill from E21 Physics Department TUM for support and with mechanical construction issues, A. Hecht for the design of the capacitance matching boxes and, together with T. Rapp for the development of the resonating circuits. G. Brandl, A.Lenz, E. Faulhaber and J. Kr\"uger from the MLZ instrument control group have adapted the NICOS control to the needs of NRSE and MIEZE. Detector support is given by the MLZ detector group, especially K. Zeitlhack and P. Wind. The CASCADE detector is also supported by M. Klein from CDT Heidelberg. Support from the neutron optics group, namely P. Link, C. Breunig, A. Ofner, E. Kahle is acknowledged. We are grateful for discussion on Larmor precession methods with T. Keller, O. Holderer and S. Pasini. We also thank A. Ostermann for supporting the McStas simulations. The students L. Spitz, W. Gottwald, M. Y. Michelis and A. Englhardt contributed in various ways.
Financial support through the BMBF project "Longitudinale Resonante Neutronen Spin-Echo Spektroskopie mit Extremer Energie-Aufl\"osung" (F\"orderkennzeichen 05K16WO6) is gratefully acknowledged. This project has received funding from the European Research Council (ERC) under the European Union’s Horizon 2020 research and innovation programme (grant agreement No 788031).

\section*{References}
%\vspace{-0.7cm}

\bibliographystyle{elsarticle-num}
\bibliography{citations}	% expects file "citations.bib"

\end{document}